 \documentclass[preprint,aps]{revtex4}

\usepackage{graphicx}
 
\begin{document}
 \newcommand{\Qed}{\rule{2.5mm}{3mm}}
 \newcommand{\balpha}{\mbox{\boldmath {$\alpha$}}}
 \draft 
%
%
\title{On the origin of families of fermions and their mass matrices -  
Approximate analyses of properties of four families within approach 
unifying spins and charges}
\author{ M. Breskvar,\\
 D. Lukman\\
and\\
N.S. Manko\v c Bor\v stnik\\
Department of Physics, University of
Ljubljana, Jadranska 19, 1000 Ljubljana,\\
}
\date{\today}
\begin{abstract} 
The approach unifying all the internal degrees of freedom - proposed by one of 
us\cite{norma92,norma93,normasuper94,norma95,norma97,pikanormaproceedings1,holgernorma00,norma01,%
pikanormaproceedings2,Portoroz03} - is offering a new way of understanding 
families of quarks and leptons. 
Spinors, namely, living in $d \;(=1+13)-$dimensional space, manifest  
in the observed $d(=1+3)$-dimensional space (at ''physical energies'')  
all the known charges of quarks and leptons (with the mass protection property of the Standard model 
- only the left handed quarks 
and leptons carry the  weak charge while the right handed ones are weak chargeless - included), 
while a part of the starting Lagrange density  in $d \;(=1+13)$ transforms the right handed 
quarks and leptons into the left handed ones, manifesting a mass term in $d=1+3$. Since a spinor  
carries two kinds of spins and interacts accordingly with two kinds of the 
spin connection fields, the approach predicts families and the corresponding Yukawa couplings.
In the paper\cite{pikanorma05} the appearance of families of quarks 
and leptons within this approach was investigated and the explicit expressions for the 
corresponding Yukawa couplings, following from the approach after some approximations and 
simplifications, presented. 
In this paper we continue investigations of this new way of presenting families of quarks and leptons 
by further analyzing properties  of mass matrices, treating quarks and leptons in an equivalent way.  
We connect free parameters of the approach with the known experimental data and investigate a  
 possibility that the fourth family of quarks and leptons appears at low enough energies to be 
observable with the new generation of accelerators.
\end{abstract}

\maketitle

\section{Introduction}
\label{introduction}

The Standard model of the electroweak and strong interactions (extended by assuming nonzero masses 
of the neutrinos) fits with around 25 parameters and constraints all the existing experimental 
data. However, it leaves  unanswered many open questions, among which are also the questions   
about the origin of the families, the Yukawa couplings of quarks and leptons and the 
corresponding Higgs mechanism.  
Understanding the mechanism for generating families, their masses and mixing matrices 
might be one of the most promising ways to 
physics beyond the Standard model. 

The approach, unifying spins and charges\cite{norma92,norma93,normasuper94,norma95,norma97,%
pikanormaproceedings1,holgernorma00,norma01,pikanormaproceedings2,Portoroz03,pikanorma05}, might by   
offering a new way of describing families, give an explanation 
about the origin of the Yukawa couplings. 

It was demonstrated in the references\cite{pikanormaproceedings1,%
norma01,pikanormaproceedings2,Portoroz03} that a left handed $SO(1,13)$ 
Weyl spinor multiplet includes, if the representation is analyzed 
in terms of the subgroups $SO(1,3)$, $SU(2)$, $SU(3)$ and the sum of the two $U(1)$'s,  all 
the spinors of the Standard model - that is the left handed $SU(2)$ doublets and the right 
handed  $SU(2)$ singlets of (with the group  $SU(3)$ charged) quarks and  (chargeless) leptons.
There are the (two kinds of)  spin connection  fields and  the vielbein field in 
$d=(1+13)-$dimensional space, 
which might  manifest - after some  
appropriate compactifications (or some other kind of making the rest of $d-4$ space 
unobservable at low energies) - in the four dimensional 
space as all the gauge fields of the known charges, as well as the Yukawa couplings.

The paper\cite{pikanorma05}  analyzes, how do terms, which lead to masses of quarks 
and leptons, appear in the approach unifying spins and charges as a part of the spin 
connection and vielbein fields. No Higgs is needed in this approach to ''dress'' 
right handed spinors with the weak charge, since the terms of the 
starting Lagrangean, which include $\gamma^0\gamma^s,$ with $s=7,8,$ do the job of a Higgs 
field.

Since we have done no analyses (yet) about the way of breaking symmetries  
of the starting group $SO(1,13)$ to 
$SO(1,7) \times U(1) \times SU(3)$ and further within our approach (except some very rough 
estimations in ref.\cite{hnrunBled02}), we do not know how might 
symmetry breaking in  the ordinary space  influence the fields, 
which determine the Yukawa couplings. We can accordingly in this investigation, by connecting 
Yukawa couplings with the experimental data, only discuss about the appearance of the 
''vacuum expectation values'' of the spin connection fields which enter into the Yukawa couplings. 

We also have no explanation yet why the second kind of the Clifford algebra objects do 
not manifest in $d=1+3$ any charges, which could appear in addition to the known ones. 

Since the generators of the Lorentz transformations and the generators of families commute, 
and since only the generators of families contribute to nondiagonal elements of mass matrices  
(which means, that the off diagonal matrix elements of quarks and leptons are strongly correlated),  
the question arises, what makes leptons so different from quarks in the 
proposed approach. Can it be that at 
some energy level they are very alike and that there are some kinds of boundary conditions 
together with the nonperturbative effects which 
lead to observable properties, or might it be that Majorana like objects, 
not taken into account in these investigations, are responsible for the observed differences?

In this paper we try to understand properties of quarks and leptons within  the approach 
unifying spins and charges treating quarks and leptons equivalently. 
Within this approach we discuss also a possibility, that 
the fourth family of quarks and leptons appears at low enough energies to be observable 
with new accelerators.

In Sect.\ref{lagrangesec} of this paper we present the action  for a Weyl spinor in 
$(1+13)$-dimensional space and the part of the  Lagrangean, which manifests at ''physical energies''   
as an effective Lagrangean, with the Yukawa mass term included. This section is a brief 
repetition of the derivations presented 
in the ref.\cite{pikanorma05}.

Sect.\ref{Lwithassumptions} repeats the explicit expressions for the four mass matrices of 
four families of quarks and leptons as derived in the  
paper\cite{pikanorma05} under several assumptions  and simplifications 
from the starting action of the approach 
unifying spins and charges. 
We study properties of the mass matrices in this assumed approximation.

In Sect.\ref{negativemasses} we discuss the problem of the appearance of negative masses 
in connection with the internal parity, defined within the presented approach. 

In Sect.\ref{improvedmatrices} we relax some of the assumptions and evaluate approximately the 
improved properties of quarks and leptons.

In Sect.\ref{discussions} we comment on the approximate predictions of our approach.

\section{Weyl spinors in $d= (1+13)$ manifesting at ''physical energies'' families of 
quarks and leptons} 
\label{lagrangesec}

We assume a  left handed Weyl spinor  in $(1+13)$-dimensional space. A spinor carries 
only the spin (no charges) and interacts accordingly with only the gauge gravitational fields 
- with spin connections and vielbeins. We assume two kinds of the Clifford algebra objects 
and allow accordingly two kinds of gauge fields\cite{norma92,norma93,normasuper94,norma95,norma97,%
pikanormaproceedings1,holgernorma00,norma01,pikanormaproceedings2,Portoroz03,pikanorma05}. 
One kind is the ordinary gauge field (gauging the Poincar\' e symmetry in $d=1+13$). 
The corresponding spin connection field appears for spinors as a gauge field of 
$S^{ab}= \frac{1}{4} (\gamma^a \gamma^b - \gamma^b \gamma^a)$, 
where $\gamma^a$ are the ordinary Dirac operators. 
The contribution of these fields to the mass matrices manifests in only the diagonal terms -  
connecting the right handed weak chargeless quarks or leptons to the left handed weak charged 
partners within one family of spinors. 

The second kind of gauge fields is in our approach responsible for families of spinors and 
couplings among families of spinors - contributing to diagonal matrix elements as well - and    
might explain the appearance of families of quarks and leptons and the Yukawa couplings of the 
Standard model of the electroweak and colour interactions. 
The corresponding spin connection fields appear for spinors as  gauge fields  
of $\tilde{S}^{ab}$ ($\tilde{S}^{ab} = \frac{1}{2} (\tilde{\gamma}^a \tilde{\gamma}^b-
\tilde{\gamma}^b \tilde{\gamma}^a)$) with $\tilde{\gamma}^a$, which are 
the Clifford algebra objects\cite{norma93,technique03}, like $\gamma^a$, but anticommute with $\gamma^a$.

Following the ref.\cite{pikanorma05} we write the action for a Weyl (massless) spinor  
in $d(=1+13)$ - dimensional space as follows\footnote{Latin indices  
$a,b,..,m,n,..,s,t,..$ denote a tangent space (a flat index),
while Greek indices $\alpha, \beta,..,\mu, \nu,.. \sigma,\tau ..$ denote an Einstein 
index (a curved index). Letters  from the beginning of both the alphabets
indicate a general index ($a,b,c,..$   and $\alpha, \beta, \gamma,.. $ ), 
from the middle of both the alphabets   
the observed dimensions $0,1,2,3$ ($m,n,..$ and $\mu,\nu,..$), indices from the bottom of 
the alphabets
indicate the compactified dimensions ($s,t,..$ and $\sigma,\tau,..$). We assume the signature 
$\eta^{ab} =
diag\{1,-1,-1,\cdots,-1\}$.
}
\begin{eqnarray}
S &=& \int \; d^dx \; {\mathcal L}
\nonumber\\
{\mathcal L} &=& \frac{1}{2} (E\bar{\psi}\gamma^a p_{0a} \psi) + h.c. = \frac{1}{2} 
(E\bar{\psi} \gamma^a f^{\alpha}{}_a p_{0\alpha}\psi) + h.c. 
\nonumber\\
p_{0\alpha} &=& p_{\alpha} - \frac{1}{2}S^{ab} \omega_{ab\alpha} - \frac{1}{2}\tilde{S}^{ab} 
\tilde{\omega}_{ab\alpha}.
\label{lagrange}
\end{eqnarray}
 
Here $f^{\alpha}{}_a$ are  vielbeins (inverted to the gauge field of the generators of translations  
$e^{a}{}_{\alpha}$, $e^{a}{}_{\alpha} f^{\alpha}{}_{b} = \delta^{a}_{b}$,
$e^{a}{}_{\alpha} f^{\beta}{}_{a} = \delta_{\alpha}{}^{\beta}$),
with $E = \det(e^{a}{}_{\alpha})$, while  
$\omega_{ab\alpha}$ and $\tilde{\omega}_{ab\alpha} $ are the two kinds of the spin connection fields, 
the gauge 
fields of $S^{ab}$ and $\tilde{S}^{ab}$, respectively, corresponding to the two kinds of the Clifford 
algebra 
objects\cite{holgernorma02,Portoroz03}, namely $\gamma^a$ and $\tilde{\gamma}^{a}$, with the 
properties
\begin{eqnarray}
\{\gamma^a,\gamma^b\}_{+} = 2\eta^{ab} =  \{\tilde{\gamma}^a,\tilde{\gamma}^b\}_{+},
\quad \{\gamma^a,\tilde{\gamma}^b\}_{+} = 0, 
\label{clifford}
\end{eqnarray}
leading to $\{ S^{ab}, \tilde{S}^{cd}\}_-=0$. 
We kindly ask the reader to learn about the properties 
of these two kinds of the Clifford algebra objects - $\gamma^a$ and $\tilde{\gamma}^a$  
and of the corresponding  $S^{ab}$ and $\tilde{S}^{ab}$ - and about our technique in the 
ref.\cite{pikanorma05} or the refs.\cite{holgernorma02,technique03}.

One Weyl spinor representation in $d=(1+13)$ with the spin as the only internal 
degree of freedom,   
manifests, if analyzed in terms of the subgroups $SO(1,3) \times
U(1) \times SU(2) \times SU(3)$ in 
four-dimensional ''physical'' space  as the ordinary ($SO(1,3)$) spinor with all the known charges 
of one family of  the left handed weak charged and the right handed weak chargeless 
quarks and leptons of the Standard model. The reader can see this analyses in the 
paper\cite{pikanorma05} (as well as in several references, like the one\cite{Portoroz03}).

We may rewrite the Lagrangean of Eq.(\ref{lagrange}) so that it manifests the usual 
$(1+3)-$dimensional spinor Lagrangean part and the term manifesting as a mass 
term\cite{pikanorma05} 
\begin{eqnarray}
{\mathcal L} &=& \bar{\psi}\gamma^{m} (p_{m}- \sum_{A,i}\; g^{A}\tau^{Ai} A^{Ai}_{m}) \psi 
+ \nonumber\\
& &  \sum_{s=7,8}\; 
\bar{\psi} \gamma^{s} p_{0s} \; \psi + {\rm the \;rest}.
\label{yukawa}
\end{eqnarray}
Index $A$ determines the charge groups ($SU(3), SU(2)$ and the two $U(1)$'s), index $i$ determines
the generators within one charge group. $\tau^{Ai}$ denote the generators of the charge groups 
\begin{eqnarray}
\tau^{Ai} = \sum_{s,t} \;c^{Ai}{ }_{st} \; S^{st},
\nonumber\\
\{\tau^{Ai}, \tau^{Bj}\}_- = i \delta^{AB} f^{Aijk} \tau^{Ak}.
\label{tau}
\end{eqnarray}
 with $s,t \in 5,6,..,14$, while $A^{Ai}_{m}, m=0,1,2,3,$ 
denote the corresponding
gauge fields (expressible in terms of $\omega_{st m}$).

We have $Y = \tau^{41} + \tau^{21}, \quad  Y' = \tau^{41} - \tau^{21},$ with $
\tau^{11}: = \frac{1}{2} ( {\mathcal S}^{58} - {\mathcal S}^{67} ),
\tau^{12}: = \frac{1}{2} ( {\mathcal S}^{57} + {\mathcal S}^{68} ),
\tau^{13}: = \frac{1}{2} ( {\mathcal S}^{56} - {\mathcal S}^{78} ),
\tau^{21}: = \frac{1}{2} ( {\mathcal S}^{56} + {\mathcal S}^{78} ),
\tau^{31}: = \frac{1}{2} ( {\mathcal S}^{9\;12} - {\mathcal S}^{10\;11} ),
\tau^{32}: = \frac{1}{2} ( {\mathcal S}^{9\;11} + {\mathcal S}^{10\;12} ),
\tau^{33}: = \frac{1}{2} ( {\mathcal S}^{9\;10} - {\mathcal S}^{11\;12} ),
\tau^{34}: = \frac{1}{2} ( {\mathcal S}^{9\;14} - {\mathcal S}^{10\;13} ),
\tau^{35}: = \frac{1}{2} ( {\mathcal S}^{9\;13} + {\mathcal S}^{10\;14} ),
\tau^{36}: = \frac{1}{2} ( {\mathcal S}^{11\;14} - {\mathcal S}^{12\;13}),
\tau^{37}: = \frac{1}{2} ( {\mathcal S}^{11\;13} + {\mathcal S}^{12\;14} ),
\tau^{38}: = \frac{1}{2\sqrt{3}} ( {\mathcal S}^{9\;10} + {\mathcal S}^{11\;12} - 2{\mathcal S}^{13\;14}),
\tau^{41}: = -\frac{1}{3}( {\mathcal S}^{9\;10} + {\mathcal S}^{11\;12} + {\mathcal S}^{13\;14}).$

The subgroups are chosen so that the gauge fields in the "physical" region agree with the
known gauge fields. If the break of symmetries in the $\tilde{S}^{ab}$ sector demonstrates 
the same symmetry after the break as in the 
$S^{ab}$ sector, then also the corresponding operators with $\tilde{\tau}^{Ai}$ should be defined.

Making several assumptions, explained in details in the ref.\cite{pikanorma05} - 
we shall repeat them bellow - 
 needed to manifest the observable phenomena (and can not yet be derived, since we do not yet know 
how the break of symmetries influences the starting Lagrangean), we are able to rewrite the mass term 
of spinors (fermions) from Eq.(\ref{yukawa}) $(\sum_{s=7,8}\; \bar{\psi} \gamma^{s} p_{0s} \; \psi$ 
(neglecting  ${\rm the \;rest})$
) by assuming that they are small in 
comparison with what we keep  at "physical energies") 
as $L_Y$, demonstrating the  Yukawa couplings of  the Standard model 
\begin{eqnarray}
{\mathcal L}_{Y} = \psi^+ \gamma^0 \;  
\{ & &\stackrel{78}{(+)} ( \sum_{y=Y,Y'}\; y A^{y}_{+} + 
\sum_{\tilde{y}=\tilde{N}^{+}_{3},\tilde{N}^{-}_{3},\tilde{\tau}^{13},\tilde{Y},\tilde{Y'}} 
\tilde{y} \tilde{A}^{\tilde{y}}_{+}\;)\; + \nonumber\\
  & & \stackrel{78}{(-)} ( \sum_{y=Y,Y'}\;y  A^{y}_{-} +  
\sum_{\tilde{y}= \tilde{N}^{+}_{3},\tilde{N}^{-}_{3},\tilde{\tau}^{13},\tilde{Y},\tilde{Y'}} 
\tilde{y} \tilde{A}^{\tilde{y}}_{-}\;) + \nonumber\\
 & & \stackrel{78}{(+)} \sum_{\{(ac)(bd) \},k,l} \; \stackrel{ac}{\tilde{(k)}} \stackrel{bd}{\tilde{(l)}}
\tilde{{A}}^{kl}_{+}((ac),(bd)) \;\;+  \nonumber\\
 & & \stackrel{78}{(-)} \sum_{\{(ac)(bd) \},k,l} \; \stackrel{ac}{\tilde{(k)}}\stackrel{bd}{\tilde{(l)}}
\tilde{{A}}^{kl}_{-}((ac),(bd))\}\psi,
\label{yukawa4tilde0}
\end{eqnarray}
with
\begin{eqnarray}
\stackrel{ab}{(k)}: = \frac{1}{2} (\gamma^a + \frac{\eta^{aa}}{ik} \gamma^8 ),\quad 
\stackrel{ab}{\tilde{(k)}}= \frac{1}{2} (\tilde{\gamma}^a + 
\frac{\eta^{aa}}{ik} \tilde{\gamma}^b )  
\label{pm}
\end{eqnarray}
and with $k=\pm 1,$ if $\eta^{aa}\eta^{bb}=1$ and $ \pm i,$ if $\eta^{aa}\eta^{bb}=-1$.
While $\stackrel{ab}{(k)}$ are expressible in terms of ordinary $\gamma^a$ and $\gamma^b$,  
$\stackrel{ab}{\tilde{(k)}}$ are expressible in terms of the second kind of the Clifford algebra objects, 
namely in terms of $\tilde{\gamma}^a$ and $\tilde{\gamma}^b$.

The Yukawa part of the starting Lagrangean (Eq.(\ref{yukawa4tilde0})) has the diagonal terms, that is the 
terms manifesting the Yukawa couplings within each family, and the off diagonal terms, determining the 
Yukawa couplings among families.

The operators, which contribute to the non diagonal terms in mass matrices, 
are superposition of $\tilde{S}^{ab}$ (times the corresponding fields $\tilde{\omega}_{abc}$)  
and can be represented  as factors of nilpotents 
\begin{eqnarray}
\stackrel{ab}{\tilde{(k)}}\stackrel{cd}{\tilde{(l)}},  
\label{lowertilde}
\end{eqnarray}
with indices $(ab)$ and $(cd)$ which belong to the Cartan subalgebra indices 
and the superposition of the  fields $\tilde{\omega}_{abc}$.
We may write accordingly
\begin{eqnarray}
  \sum_{(a,b) } -\frac{1}{2} \stackrel{78}{(\pm)}\tilde{S}^{ab} \tilde{\omega}_{ab\pm} =
- \sum_{(ac),(bd), \;  k,l}\stackrel{78}{(\pm)}\stackrel{ac}{\tilde{(k)}}\stackrel{bd}{\tilde{(l)}} 
\; \tilde{A}^{kl}_{\pm} ((ac),(bd)),  
\label{lowertildeL}
\end{eqnarray}
where the pair $(a,b)$ in the first sum runs over all the  indices, which do not characterize  the Cartan 
subalgebra, with $ a,b = 0,\dots, 8$,  while the two pairs $(ac)$ and $(bd)$ denote only the Cartan 
subalgebra pairs
 (for SO(1,7) we only have the pairs $(03), (12)$; $(03), (56)$ ;$(03), (78)$;
$(12),(56)$; $(12), (78)$; $(56),(78)$) ; $k$ and $l$ run over four 
possible values so that $k=\pm i$, if $(ac) = (03)$ 
and $k=\pm 1$ in all other cases, while $l=\pm 1$.
The fields  $\tilde{A}^{kl}_{\pm} ((ac),(bd))$ can then 
be expressed by $\tilde{\omega}_{ab \pm}$ as follows 
\begin{eqnarray}
\tilde{A}^{++}_{\pm} ((ab),(cd)) &=& -\frac{i}{2} (\tilde{\omega}_{ac\pm} -\frac{i}{r} \tilde{\omega}_{bc\pm} 
-i \tilde{\omega}_{ad\pm} -\frac{1}{r} \tilde{\omega}_{bd\pm} ), \nonumber\\
\tilde{A}^{--}_{\pm} ((ab),(cd)) &=& -\frac{i}{2} (\tilde{\omega}_{ac\pm} +\frac{i}{r} \tilde{\omega}_{bc\pm} 
+i \tilde{\omega}_{ad\pm} -\frac{1}{r} \tilde{\omega}_{bd\pm} ),\nonumber\\
\tilde{A}^{-+}_{\pm} ((ab),(cd)) &=& -\frac{i}{2} (\tilde{\omega}_{ac\pm} + \frac{i}{r} \tilde{\omega}_{bc\pm} 
-i  \tilde{\omega}_{ad\pm} +\frac{1}{r} \tilde{\omega}_{bd\pm} ), \nonumber\\
\tilde{A}^{+-}_{\pm} ((ab),(cd)) &=& -\frac{i}{2} (\tilde{\omega}_{ac\pm} - \frac{i}{r} \tilde{\omega}_{bc\pm} 
+i  \tilde{\omega}_{ad\pm} +\frac{1}{r} \tilde{\omega}_{bd\pm} ),
\label{Awithomega}
\end{eqnarray}
with $r=i$, if $(ab) = (03)$ and $r=1$ otherwise.
 We simplify the index $kl$ in the exponent 
of the fields $\tilde{A}^{kl}{}_{\pm} ((ac),(bd))$ to $\pm $, omitting $i$. 

We must point out that a way of breaking  any of the two symmetries - the Poincar\' e one and 
the symmetry  determined by the generators $\tilde{S}^{ab}$ in $d=1+13$ - 
strongly influences the Yukawa couplings of 
Eq.(\ref{yukawa4tilde0}), relating  the parameters $\tilde{\omega}_{abc}$. 
Not necessarily any break of  the Poincar\' e   symmetry 
 influences  the break of the other symmetry and opposite. Although we expect that it does. 
 Accordingly 
the coefficients $c^{Ai}{}_{ab}$ determining 
the operators $\tau^{Ai}$ in Eq.(\ref{tau}) and the coefficients  
$\tilde{c}^{\tilde{A}i}{}_{ab}$ 
determining the operators $\tilde{\tau}^{\tilde{A}i}$ in the relations 
\begin{eqnarray}
\tilde{\tau}^{\tilde{A}i} = \sum_{a,b} \;\tilde{c}^{\tilde{A}i}{ }_{ab} \; \tilde{S}^{ab},
\nonumber\\
\{\tilde{\tau}^{\tilde{A}i}, \tilde{\tau}^{\tilde{B}j}\}_- = i \delta^{\tilde{A}\tilde{B}} 
\tilde{f}^{\tilde{A}ijk} 
\tilde{\tau}^{\tilde{A}k}
\label{tildetau}
\end{eqnarray}
might even  not be correlated. If correlated (through 
boundary conditions, for example) the break of symmetries might cause that off diagonal 
matrix elements of Yukawa couplings distinguish between quarks and leptons.

We made, when deriving the mass matrices of quarks and leptons from the approach 
unifying spins and charges, several assumptions, approximations and simplifications in order 
to be able to make at the end some rough  predictions about the properties of the 
families of quarks and leptons:

i.  The break of symmetries of the group $SO(1,13)$ (the  Poincar\' e group in 
$d=1+13$)  into $SO(1,7)\times SU(3)\times U(1)$ occurs 
in a way that only 
massless spinors in $d=1+7$ with the charge $ SU(3)\times U(1)$ survive, and yet the two $U(1)$ charges, 
following from $SO(6)$ and $SO(1,7)$, respectively, are related. (Our work on the compactification 
of a massless spinor in $d=1+5$ into   $d=1+3$ and a finite disk gives us some hope that such 
an assumption might be justified\cite{holgernorma05}.) 
The requirement that the terms with $S^{5a}\omega_{5ab}$ and $S^{6a}\omega_{6ab}$ 
do not contribute to the mass term, assures that the charge   
$Q= \tau^{41} + S^{56}$ is conserved at low energies. 

ii. The break of symmetries influences  both, the (Poincar\' e) symmetry 
described by $S^{ab}$ and the symmetries described by $\tilde{S}^{ab}$, and in a way that  
 there are no terms, which would  transform  
$\stackrel{{56}}{\tilde{(+)}}$ into $\stackrel{{56}}{\tilde{[+]}}$. This assumption  was made that 
at "low energies" only four families have to be treated and 
can be explained by a break of the symmetry $SO(1,7)$ into $SO(1+5)\times U(1)$ in the $\tilde{S}^{ab}$ 
sector so that all the contributions of the type $\tilde{S}^{5a}\tilde{\omega}_{5ab}$ and 
$\tilde{S}^{6a}\tilde{\omega}_{6ab}$ are equal to zero. We also assume that the terms which 
include components $p_s, s=5,..,14$, of the momentum $p^a$ do not 
contribute  to the mass matrices. We keep in mind that  any further break of symmetries 
strongly influences the relations among 
$\tilde{\omega}_{abc}$, appearing in the paper \cite{pikanorma05} as ''vacuum expectation values'' 
in mass matrices, so that predictions in Sect.\ref{numerical} strongly depend on the way of breaking. 
 
iv. We  make estimations on a ''tree level''.  

v. We assume the mass matrices to  be real and symmetric (expecting that complexity and nonsymmetric 
properties will not influence considerably masses and 
mixing matrices of quarks and leptons).


\section{Four families of quarks and leptons}
\label{Lwithassumptions}

Taking into account the assumptions, presented in Sect.\ref{lagrangesec}, we end up with four 
families of quarks and leptons 
\begin{eqnarray}
I.\;& & \stackrel{03}{(+i)} \stackrel{12}{(+)} |\stackrel{56}{(+)} \stackrel{78}{(+)}||...\nonumber\\ 
II.\;& &\stackrel{03}{[+i]} \stackrel{12}{[+]} |\stackrel{56}{(+)} \stackrel{78}{(+)}||... \nonumber\\
III.& & \stackrel{03}{[+i]} \stackrel{12}{(+)} |\stackrel{56}{(+)} \stackrel{78}{[+]}||... \nonumber\\
IV. & & \stackrel{03}{(+i)} \stackrel{12}{[+]} |\stackrel{56}{(+)} \stackrel{78}{[+]}||... .
\label{fourfamilies}
\end{eqnarray}
The Yukawa couplings for these four families are for $u$-quarks and neutrinos  
presented on Table~\ref{TableI}, where 
$\alpha$ stays for $u$-quarks and neutrinos.
\begin{table}
\begin{center}
\begin{tabular}{|r||c|c|c|c|}
\hline
$\alpha$&$ I_{R} $&$ II_{R} $&$ III_{R} $&$ IV_{R}$\\
\hline\hline
&&&& \\
$I_{L}$   & $ A^I_{\alpha}  $ & $ \tilde{A}^{++}_{\alpha} ((03),(12))= $ & $  \tilde{A}^{++}_{\alpha} ((03),(78)) =$  &
$ -  \tilde{A}^{++}_{\alpha} ((12),(78)) = $ \\
$$&$$ &$ \frac{1}{2}(\tilde{\omega}_{327} +\tilde{\omega}_{018})$&$ \frac{1}{2}(\tilde{\omega}_{387} +\tilde{\omega}_{078}) $&$
 \frac{1}{2}(\tilde{\omega}_{277} +\tilde{\omega}_{187})$ \\
&&&& \\
\hline
&&&&\\
$II_{L}$  & $ \tilde{A}^{--}_{\alpha} ((03),(12))= $ & $ A^{II}_{\alpha}= $ & $  \tilde{A}^{-+}_{\alpha} ((12),(78)) = $ &
$ -  \tilde{A}^{-+}_{\alpha} ((03),(78)) = $ \\
$$&$ \frac{1}{2}(\tilde{\omega}_{327} +\tilde{\omega}_{018})$&$A^{I}_{\alpha} +  (\tilde{\omega}_{127} - \tilde{\omega}_{038})$
&$ -\frac{1}{2}(\tilde{\omega}_{277} -\tilde{\omega}_{187}) $&$  \frac{1}{2}(\tilde{\omega}_{387} - \tilde{\omega}_{078})$ \\
\hline 
&&&& \\
$III_{L}$ & $  \tilde{A}^{--}_{\alpha} ((03),(78)) =$ & $- \tilde{A}^{+-}_{\alpha} ((12),(78))= $ & $ A^{III}_{\alpha}=$ &
$   \tilde{A}^{-+}_{\alpha} ((03),(12)) =$ \\
$$&$  \frac{1}{2}(\tilde{\omega}_{387} +\tilde{\omega}_{078})$&$ -\frac{1}{2}(\tilde{\omega}_{277} -\tilde{\omega}_{187})$
&$A^{I}_{\alpha} +  (\tilde{\omega}_{787} - \tilde{\omega}_{038})$&$ -\frac{1}{2}(\tilde{\omega}_{327} -\tilde{\omega}_{018}) $\\
&&&& \\
\hline 
&&&& \\
$IV_{L}$  & $ \tilde{A}^{--}_{\alpha} ((12),(78)) =$ & $- \tilde{A}^{+-}_{\alpha} ((03),(78)) = $ & 
$ \tilde{A}^{+-}_{\alpha} ((03),(12))$ & $ A^{IV}_{\alpha} = $ \\
$$&$ \frac{1}{2}(\tilde{\omega}_{277} +\tilde{\omega}_{187})$&$ \frac{1}{2}(\tilde{\omega}_{387} -\tilde{\omega}_{078})$
&$ -\frac{1}{2}(\tilde{\omega}_{327} -\tilde{\omega}_{018}) $&$A^{I}_{\alpha} +  (\tilde{\omega}_{127} + \tilde{\omega}_{787})$\\
&&&& \\
\hline\hline
\end{tabular}
\end{center}
\caption{\label{TableI}%
The mass matrix of four families of  $u$-quarks and neutrinos, obtained within the approach 
unifying spins and charges under the assumptions i.-v. in ref.\cite{pikanorma05}. 
According to Eq.(\ref{yukawadiagonal}) and Table~\ref{TableI} and~\ref{TableII}
there are 13 free parameters, expressed in terms
of the fields $A^{I}_{\alpha}$ and $\tilde{\omega}_{abc}$, which accordingly determine (due to assumptions 
i.-v.) all the properties of the  four families of the two types of quarks and the two types of 
leptons.} 
\end{table}

The corresponding mass matrix for the $d$-quarks and the electrons is presented on Table~\ref{TableII}, 
where $\beta$ stays for $d$-quarks and electrons.
\begin{table}
\begin{center}
\begin{tabular}{|r||c|c|c|c|}
\hline
$\beta$&$ I_{R} $&$ II_{R} $&$ III_{R} $&$ IV_{R}$\\
\hline\hline
&&&& \\
$I_{L}$   & $ A^I_{\beta}  $ & $ \tilde{A}^{++}_{\beta} ((03),(12))= $ & $  - \tilde{A}^{++}_{\beta} ((03),(78)) =$  &
$   \tilde{A}^{++}_{\beta} ((12),(78)) = $ \\
$$&$$ &$ \frac{1}{2}(\tilde{\omega}_{327} - \tilde{\omega}_{018})$&$ -\frac{1}{2}(\tilde{\omega}_{387} - 
\tilde{\omega}_{078}) $&$  -\frac{1}{2}(\tilde{\omega}_{277} +\tilde{\omega}_{187})$ \\
&&&& \\
\hline
&&&&\\
$II_{L}$  & $ \tilde{A}^{--}_{\beta} ((03),(12))= $ & $ A^{II}_{\beta}= $ & $  -\tilde{A}^{-+}_{\beta} ((12),(78)) = $ &
$  \tilde{A}^{-+}_{\beta} ((03),(78)) = $ \\
$$&$ \frac{1}{2}(\tilde{\omega}_{327} -\tilde{\omega}_{018})$&$A^{I}_{\beta} +  (\tilde{\omega}_{127} + \tilde{\omega}_{038})$
&$ \frac{1}{2}(\tilde{\omega}_{277} -\tilde{\omega}_{187}) $&$  -\frac{1}{2}(\tilde{\omega}_{387} +\tilde{\omega}_{078})$ \\
\hline 
&&&& \\
$III_{L}$ & $  - \tilde{A}^{--}_{\beta} ((03),(78)) =$ & $ \tilde{A}^{+-}_{\beta} ((12),(78))= $ & $ A^{III}_{\beta}=$ &
$   \tilde{A}^{-+}_{\beta} ((03),(12)) =$ \\
$$&$ -\frac{1}{2}(\tilde{\omega}_{387} -\tilde{\omega}_{078})$&$ \frac{1}{2}(\tilde{\omega}_{277} -\tilde{\omega}_{187})$
&$A^{I}_{\beta} +  (\tilde{\omega}_{787} + \tilde{\omega}_{038})$&$ -\frac{1}{2}(\tilde{\omega}_{018} +\tilde{\omega}_{327}) $\\
&&&& \\
\hline 
&&&& \\
$IV_{L}$  & $ -\tilde{A}^{--}_{\beta} ((12),(78)) =$ & $ \tilde{A}^{+-}_{\beta} ((03),(78)) = $ & 
$ \tilde{A}^{+-}_{\beta} ((03),(12))$ & $ A^{IV}_{\beta} $ \\
$$&$ -\frac{1}{2}(\tilde{\omega}_{277} +\tilde{\omega}_{187})$&$ -\frac{1}{2}(\tilde{\omega}_{387} +\tilde{\omega}_{078})$
&$ -\frac{1}{2}(\tilde{\omega}_{018} +\tilde{\omega}_{327}) $&$A^{I}_{\beta} +  (\tilde{\omega}_{127} + \tilde{\omega}_{787})$\\
&&&& \\
\hline\hline
\end{tabular}
\end{center}
\caption{\label{TableII}%
The mass matrix of four families of the $d$-quarks and electrons, $\beta$ stays for the $d$-quarks 
and the electrons. Comments are the same as on Table~\ref{TableI}.}
\end{table}

The explicit form of the diagonal matrix elements for the above choice of assumptions in terms of 
$\omega_{abc}$'s and $A^{y}_{\pm}, y = Y$ and $Y'$, 
$\tilde{\omega}_{abc}$  and $\tilde{A}^{41}_{\pm}$ is  as follows
\begin{eqnarray}
A^{I}_{u} &=& \frac{2}{3} A^{Y}_{-} - \frac{1}{3} A^{Y'}_{-}  + \tilde{\omega}^{I}_{-},\quad \;\;\;
A^{I}_{\nu}=                                    - A^{Y'}_{-}  + \tilde{\omega}^{I}_{-},\nonumber\\ 
A^{I}_{d} &=&-\frac{1}{3} A^{Y}_{+} + \frac{2}{3} A^{Y'}_{+}  + \tilde{\omega}^{I}_{+},\;\;\; \;
A^{I}_{e}  =            - A^{Y}_{+} +                           \tilde{\omega}^{I}_{+},\nonumber\\
A^{II}_{\alpha}  &=&  A^{I}_{\alpha} + (i \tilde{\omega}_{03-} + \tilde{\omega}_{12 -}), \;\;\;
A^{II}_{\beta} =  A^{I}_{\beta} + (i \tilde{\omega}_{03+} + \tilde{\omega}_{12 +}), \nonumber\\
A^{III}_{\alpha} &= & A^{I}_{\alpha} + (i \tilde{\omega}_{03-} + \tilde{\omega}_{78 -}), \;\;
A^{III}_{\beta} =  A^{I}_{\beta} + (i \tilde{\omega}_{03+} + \tilde{\omega}_{78 +}), \nonumber\\
A^{IV}_{\alpha} &= & A^{I}_{\alpha} + ( \tilde{\omega}_{12-} + \tilde{\omega}_{78 -}), \quad
A^{IV}_{\beta} =  A^{I}_{\beta} + ( \tilde{\omega}_{12+} + \tilde{\omega}_{78 +}), \nonumber\\
\label{yukawadiagonal}
\end{eqnarray}
with $\alpha= u,\nu$, $\beta=d,e $ and $-\tilde{\omega}^{I}{}_{\pm} = 
\frac{1}{2} (i \tilde{\omega}_{03 \pm} + \tilde{\omega}_{12 \pm} + 
\tilde{\omega}_{56 \pm} + \tilde{\omega}_{78 \pm}
+ \frac{1}{3} \tilde{A}^{41}_{\pm})$. The assumption that all the matrix elements  are real relates 
$\tilde{\omega}^{I}{}_{+}= \frac{1}{2} \tilde{\omega}_{03 8} + \tilde{\omega}, 
\tilde{\omega}^{I}{}_{-}= -\frac{1}{2} \tilde{\omega}_{03 8} + \tilde{\omega},$ where 
$\tilde{\omega}$ is (in case that breaking of symmetries does not influence quarks and leptons 
differently) one common parameter.

If the break of symmetries does not influence the quarks and the leptons in a different way, then 
under the assumptions i.-v. the off diagonal matrix elements of mass matrices for quarks 
are the same as for the corresponding leptons (the off diagonal matrix elements  of the 
$u$-quarks and the neutrinos are the same,  and the off diagonal matrix elements for the
$d$-quarks and the electrons are the same) and since the diagonal matrix elements differ 
only in a constant times a unit matrix, the predicted mixing matrices of 
the quarks and the leptons  would be the same. 

{\em We must ask ourselves at this stage: Can we find any way of breaking symmetries 
 - allowing any very special boundary conditions - which would lead to 
 so different  properties of quarks and leptons 
as observed or we must take the Majorana like degrees of freedom into account?}

In this paper, we are not yet able to answer this question. We can only make some estimates trying 
to  learn from the approach unifying spins and charges 
about possible explanations 
for the properties of quarks and leptons.

We proceed by relating the experimental data and the mass matrices from the approach.  
Knowing  from the experimental data that 
the first two families of quarks and leptons 
are much lighter than the third one, while in the refs.\cite{okun,okunmaltoni,okunbulatov} 
 the authors, analyzing the  
experimental data, conclude that the experimental data do not forbid  masses of the fourth family of 
quarks to be between 
$200$ GeV and $300$ GeV,  of the fourth electron to be around $100 GeV$ and of the fourth neutrino
to be at around $50 $ GeV 
we make one more assumption, 
which seems quite reasonable also 
from the point of view of the measured matrix elements of the mixing matrix for quarks. 
Namely, {\em we assume  
that the mass matrices of the four families of quarks and leptons are diagonalizable in 
two steps, so that the first diagonalization transforms 
them into block-diagonal matrices with two  $2\times 2$ sub-matrices.}  
This assumption, which   means, 
that a real and symmetric 
$4 \times 4$ matrix is diagonalizable by only three rather than six angles,  
 simplifies considerably further studies, making conclusions very transparent. 
 Such a property of  mass matrices could  be a consequence of an approximate break of symmetry 
 in the $\tilde{S}^{ab}$ sector from  $SO(1,5)$ to  $SU(2)\times SU(2)\times U(1)$, which makes, 
 for example, all the terms $\tilde{S}^{7a} \tilde{\omega}_{7ab}$ and $\tilde{S}^{8a} 
 \tilde{\omega}_{8ab}$ contributing small terms to mass matrices.  
 The exact break of this type  makes that the lower two 
 families  completely decouple from the higher two.
(Similarly we have required, in order to end up with only four rather than eight families, 
 that $SO(1,7)$ breaks to $SO(1,5) \times U(1)$ so that  
 all $\tilde{S}^{5a}\;\tilde{\omega}_{5a \pm}$ 
 and $\tilde{S}^{6a}\;\tilde{\omega}_{6a \pm}$ contribute nothing to mass matrices.) 
 While the exact break of $SO(1,5)$ to  $SU(3)\times U(1)$ makes that the fourth family decouples 
 from the first three.

It is easy to prove that a $4\times 4$ matrix is diagonalizable in two steps only  
 if it has a  structure
\begin{equation}\label{deggen}
          \left(\begin{array}{cc}
                  A   &  B\nonumber\\
                  B   &  C=A+k B \nonumber\\
                    \end{array}
                \right).
\end{equation}
Since $A$ and $C$ are, as assumed on Table~\ref{TableI} and Table~\ref{TableII}, symmetric  $2 \times 2$ matrices, 
so must then be also $B$. The parameter $k$ is assumed to be an unknown  number.

The assumption (\ref{deggen})  requires: i. $\tilde{\omega}_{277}= 0, \;
\tilde{\omega}_{327}= - \frac{k}{2}\tilde{\omega}_{187}, \; \tilde{\omega}_{787} = 
 \frac{k}{2}\tilde{\omega}_{387}, \;\tilde{\omega}_{038}=- \frac{k}{2}\tilde{\omega}_{078}$, 
and ii. $k_u= -k_d$ and $k_{\nu}= - k_{e}$, where $k_u$ and $k_{\nu}$ 
will be taken as two independent parameters. (If $k=0$ in Eq. (\ref{deggen}), the angle of rotation is
$45^\circ$ - then, if also all the $2 \times 2$ matrices would have the same structure  
(namely equal diagonal and equal nondiagonal elements),  the 
corresponding mixing matrices for quarks and leptons would be the 
identity.
)

We shall present in what follows some simple relations which  
demonstrate transparently  properties of mass matrices.
After the one step diagonalization determined by the angle of rotation
\begin{eqnarray}
\tan \varphi_{\alpha} &=&  \pm(\sqrt{1+ (\frac{k}{2})^2 } \pm \frac{k}{2}),\quad 
\tan \varphi_{\beta} =  \pm(\sqrt{1+ (\frac{k}{2})^2 } \mp \frac{k}{2}),\nonumber\\ 
{\rm with}\;&&
\tan (\varphi_{\alpha}-\varphi_{\beta}) = \pm \frac{k}{2},\; ({\rm or} \: \pm \frac{2}{k}) 
\label{firstangle}
\end{eqnarray}
we end up with two  by diagonal matrices, with $k=k_u$ for quarks and $k=k_{\nu}$ for leptons, 
while $\alpha$ concerns the $u$-quarks and 
$\nu$,  and $\beta$ the $d$-quarks  and electrons.

The first by diagonal mass matrix of the $u$-quarks ($\alpha = u$) and neutrinos ($\alpha = \nu$) is 
as follows 
\begin{equation}\label{aunuIIa}
 {\bf A^{a}_{\alpha}} = \left(\begin{array}{cc}
 a_{\alpha} 
 ,& \frac{1}{2}(\tilde{\omega}_{018} - 
 \sqrt{1 + (\frac{k_{\alpha}}{2})^2}\;\tilde{\omega}_{187})\nonumber\\
  \frac{1}{2}(\tilde{\omega}_{018} - 
 \sqrt{1 + (\frac{k_{\alpha}}{2})^2}\;\tilde{\omega}_{187}), &  a_{\alpha} + \tilde{\omega}_{127}  + 
   \sqrt{1 + (\frac{k_{\alpha}}{2})^2}\;\tilde{\omega}_{078}\nonumber\\
                    \end{array}
                \right),
\end{equation}
with $a_{u}= \frac{2}{3}A^Y - \frac{1}{3} A^{Y'} + \tilde{\omega} - \frac{1}{2} \tilde{\omega}_{038}
+ \frac{1}{2}( \frac{k_u}{2} - \sqrt{1 + (\frac{k_u}{2})^2}) 
 (\tilde{\omega}_{078} + \tilde{\omega}_{387})$ and $a_{\nu}= 
 - A^{Y'} + \tilde{\omega} - \frac{1}{2} \tilde{\omega}_{038}
+ \frac{1}{2}( \frac{k_{\nu}}{2} - \sqrt{1 + (\frac{k_{\nu}}{2})^2}) 
 (\tilde{\omega}_{078} + \tilde{\omega}_{387})$. 
 The mass matrix for the second two families of $u$-quarks ($\alpha = u$) and neutrinos 
 ($\alpha = \nu$) is equal to  
\begin{equation}\label{aunuIIb}
 {\bf A^{b}_{\alpha}} = \left(\begin{array}{cc}
 a_{\alpha} + \sqrt{1 + (\frac{k_{\alpha}}{2})^2} 
 \;(\tilde{\omega}_{078} + \tilde{\omega}_{387})
 ,& \frac{1}{2}(\tilde{\omega}_{018} + 
 \sqrt{1 + (\frac{k_{\alpha}}{2})^2}\;\tilde{\omega}_{187})\nonumber\\
  \frac{1}{2}(\tilde{\omega}_{018} +  
 \sqrt{1 + (\frac{k_{\alpha}}{2})^2}\;\tilde{\omega}_{187}), &  a_{\alpha} + \tilde{\omega}_{127}  + 
   \sqrt{1 + (\frac{k_{\alpha}}{2})^2}\;\tilde{\omega}_{387}\nonumber\\
                    \end{array}
                \right).
\end{equation}

Accordingly we find for the first two families of $d$-quarks  ($\beta = d$) and 
electrons ($\beta = e $)
\begin{equation}\label{adeIIa}
 {\bf A^{a}_{\beta}} = \left(\begin{array}{cc}
 a_{\beta} 
 ,& -\frac{1}{2}(\tilde{\omega}_{018} - 
 \sqrt{1 + (\frac{k_{\alpha}}{2})^2}\;\tilde{\omega}_{187})\nonumber\\
  -\frac{1}{2}(\tilde{\omega}_{018} - 
 \sqrt{1 + (\frac{k_{\alpha}}{2})^2}\;\tilde{\omega}_{187}), &  a_{\beta} + \tilde{\omega}_{127}  + 
   \sqrt{1 + (\frac{k_{\alpha}}{2})^2}\;\tilde{\omega}_{078}\nonumber\\
                    \end{array}
                \right),
\end{equation}
with $a_{d}= - \frac{1}{3}A^Y + \frac{2}{3} A^{Y'} + \tilde{\omega} + \frac{1}{2} \tilde{\omega}_{038}
- \frac{1}{2}( \frac{k_u}{2} + \sqrt{1 + (\frac{k_u}{2})^2}) 
 (\tilde{\omega}_{078} - \tilde{\omega}_{387})$ and $a_{e}= 
 - A^{Y} + \tilde{\omega} + \frac{1}{2} \tilde{\omega}_{038}
- \frac{1}{2}( \frac{k_{\nu}}{2} + \sqrt{1 + (\frac{k_{\nu}}{2})^2}) 
 (\tilde{\omega}_{078} - \tilde{\omega}_{387})$.  $k_{\alpha}$ in (\ref{adeIIa}) is $k_u$ 
 for $d$-quarks and $k_{\nu}$ for electrons.
For the second two families of $d$-quarks ($\beta = d $) and electrons ($\beta = e$) it follows  
\begin{equation}\label{adeIIb}
 {\bf A^{b}_{\beta}} = \left(\begin{array}{cc}
 a_{\beta} + \sqrt{1 + (\frac{k_{\alpha}}{2})^2} \;
 (\tilde{\omega}_{078} - \tilde{\omega}_{387})
 ,& -\frac{1}{2}(\tilde{\omega}_{018} +  
 \sqrt{1 + (\frac{k_{\alpha}}{2})^2}\;\tilde{\omega}_{187})\nonumber\\
  -\frac{1}{2}(\tilde{\omega}_{018} +  
 \sqrt{1 + (\frac{k_{\alpha}}{2})^2}\;\tilde{\omega}_{187}), &  a_{\beta} + \tilde{\omega}_{127}  - 
   \sqrt{1 + (\frac{k_{\alpha}}{2})^2}\;\tilde{\omega}_{387}\nonumber\\
                    \end{array}
                \right).
\end{equation}
Again, $k_{\alpha}$ in (\ref{adeIIb}) is $k_u$ for $d$-quarks and
 $k_{\nu}$ for electrons.

There are three angles, which in the two step orthogonal transformations rotate each mass matrix 
into a diagonal one. The angles of rotations for $u-$quarks  and $d-$quarks, and accordingly 
for neutrinos and electrons, 
 are related as seen from Eq.(\ref{firstangle}) for the first step rotation. It follows namely 
 that $\tan \varphi_{\alpha} = \tan^{-1} \varphi_{\beta}$ 
 and accordingly 
 \begin{eqnarray}
 \label{phirelated}
  \varphi_{\alpha} &=& \frac{\pi}{2}  -\varphi_{\beta}, \quad \varphi_{\alpha} = \frac{\pi}{4}-
  \frac{\varphi}{2}, \quad \varphi_{\beta} = \frac{\pi}{4}+   \frac{\varphi}{2},\nonumber\\
  \quad {\rm with}\;&& \varphi = \varphi_{\alpha} -\varphi_{\beta}.
  \end{eqnarray}
Similarly also the two angles of rotations of the two by two diagonal matrices are related.
Reminding the reader that in the unitary transformations ($S^{\dagger} S=I$)  
the  trace and the determinant are among the invariants, while the angle 
of rotation, which diagonalizes $2$ by $2$ matrices  (of the type (\ref{deggen})), and the 
values of the diagonal matrices are related as follows
\begin{eqnarray}
\tan \Phi &=& (\sqrt{1+(\frac{C-A}{2B})^2}\mp \frac{C-A}{2B}),\nonumber\\
\lambda_{1,2} &=& \frac{1}{2}((C+A) \pm \sqrt{(C-A)^2 + (2B)^2}),
\label{solutions}
\end{eqnarray}
where for $A,B,C$ the corresponding matrix elements from Eqs.(\ref{aunuIIa},%
\ref{aunuIIb},\ref{adeIIa},\ref{adeIIb})  must be taken, one   
  easily finds that  
\begin{eqnarray}
\label{angle}
{}^a \eta_{\alpha}= -{}^a \eta_{\beta},\quad {}^b \eta_{\alpha}= -{}^b \eta_{\beta},
 \quad \alpha = u,\nu,\; \beta = d, e,
\end{eqnarray}
where index $a$ denotes the first two families and $b$ the second two families of 
either quarks ($\alpha =u,\; \beta = d$) and leptons ($\alpha =\nu,\; \beta = e$) and 
$\eta = \frac{C-A}{2B}$. 
One then finds the relations, equivalent to those of  Eq.(\ref{phirelated}) 
 \begin{eqnarray}
 \label{abphirelated}
  {}^{a,b}\varphi_{\alpha} &=& \frac{\pi}{2}  - {}^{a,b}\varphi_{\beta}, 
  \quad {}^{a,b}\varphi_{\alpha} = \frac{\pi}{4}-
  \frac{{}^{a,b}\varphi}{2}, \quad {}^{ab}\varphi_{\beta} = \frac{\pi}{4}+   
  \frac{{}^{a,b}\varphi}{2},\nonumber\\
  \quad {\rm with}\;&& {}^{a,b}\varphi = {}^{a,b}\varphi_{\alpha} -{}^{a,b}\varphi_{\beta}.
  \end{eqnarray}
We further find 
\begin{eqnarray}
\label{consequences}
        |m_{u_2} - m_{u_1}| = |m_{d_2} - m_{d_1}|,& & \;
       |m_{u_4} - m_{u_3}|  =  |m_{d_4} - m_{d_3}|,\nonumber\\
|m_{\nu_{2}} - m_{\nu_{1}}| = |m_{e_2} -m_{e_1}|, & & \;
|m_{\nu_4} -m_{\nu_{3}}| = |m_{e_4} -m_{e_3}|,\nonumber\\
|(m_{u_4} + m_{u_3}) - (m_{u_2} + m_{u_1})| &=& |(m_{d_4} + 
m_{d_3}) - (m_{d_2} + m_{d_1})|, \nonumber\\ 
|(m_{\nu_4} + m_{\nu_{3}}) - (m_{\nu_{2}} + m_{\nu_1})| &=&  
|(m_{e_4} + m_{e_3}) - (m_{e_2} + m_{e_1})|,\nonumber\\
|m_{u_4} + m_{u_3}| & \approx & 2\;\sqrt{1+(\frac{k_u}{2})^2} \;\tilde{\omega}_{387} 
 \approx |m_{d_4} + m_{d_3}|, \nonumber\\ 
 |m_{\nu_4} + m_{\nu_{3}}| &\approx & 2\; \sqrt{1+(\frac{k_{\nu}}{2})^2} \;\tilde{\omega}_{387} 
 \approx |m_{e_4} + m_{e_3}|.  
\end{eqnarray}
We take the absolute values of the sums and the differences, since 
whenever an eigenvalue $\lambda_{1,2} $ (Eq.\ref{solutions}) appears to be negative,   
an appropriate change of a phase of the corresponding  state transforms the negative value into the 
positive one by changing simultaneously the internal parity of the particular state, 
as it will be discussed in Sect.\ref{negativemasses}. 

The above relations (\ref{consequences}) do not agree with the  experimental data. 
We can only accept them as a very rough estimation - after making so  
many assumptions and approximations - in the limit if masses of the  fourth family 
are much higher than the mass $m_{t}$, knowing   
 that $m_{t} $ is more than $30$ times larger than the mass $m_{b}$.

It is obvious that we made on the way of deriving properties of quarks and leptons from the starting 
action (Eq.\ref{lagrange}) (for a spinor carrying only two kinds of the spin degrees of freedom - 
no charge - and interacting with only the gauge fields of the corresponding groups, which  
leads to the Yukawa mass matrices (Eq.\ref{yukawa}) without assuming Higgs)  
so many assumptions, simplifications and approximations, that we lose the predictive power. 

We did not (could not yet - to do this is a huge project, which we do have in mind) 
take into account influences of possible breaks of symmetries, 
which would certainly 
bring  relations among $\tilde{\omega}_{abc}$ fields, but could also - due to some boundary 
conditions or some other effects - change the relations  among  $\tilde{\omega}_{abc}$ 
fields for quarks and leptons. One could also expect that  possible nonperturbative 
effects might be a very strong reason for the differences among properties of 
observed fermions. 
In sect.\ref{improvedmatrices} we shall try to simulate these effects in a very rough way - 
just by assuming that the off diagonal matrix elements might be different for different species of 
fermions while  keeping the relations among the angles of 
the orthogonal rotations from Eq.(\ref{firstangle},\ref{phirelated},\ref{angle},\ref{abphirelated}).

 
\section{Negative masses and parity of states}
\label{negativemasses}

We have mentioned in the previous  section that after the diagonalization of mass matrices 
of quarks and leptons, masses of either positive or negative sign can appear.

Let us first recognize that while the starting Lagrange density for spinors (Eq.\ref{lagrange}) 
commutes with the operator of handedness in $d(=1+13)$-dimensional space $\Gamma^{(1,13)} $ 
($\Gamma^{(1,13)} \;= i 2^{7} 
\; S^{03} S^{12} S^{56} \cdots S^{13 \; 14} $), it does not commute with the operator of handedness 
in $d(=1+3)$-dimensional space $\Gamma^{(1,3)} $ ($\Gamma^{(1,3)}\;=  - i 2^2 S^{03} S^{12}$). 
Accordingly also the term, which 
manifests at "physical energies" as the mass term $ \hat{m}$ ($ \gamma^0 \hat{m}=
\gamma^0 \;  
\{ \stackrel{78}{(+)} ( \sum_{y=Y,Y'}\; y A^{y}_{+} + 
\sum_{\tilde{y}=\tilde{N}^{+}_{3},\tilde{N}^{-}_{3},\tilde{\tau}^{13},\tilde{Y},\tilde{Y'}} 
\tilde{y} \tilde{A}^{\tilde{y}}_{+}\;)\; + 
   \stackrel{78}{(-)} ( \sum_{y=Y,Y'}\;y  A^{y}_{-} +  
\sum_{\tilde{y}= \tilde{N}^{+}_{3},\tilde{N}^{-}_{3},\tilde{\tau}^{13},\tilde{Y},\tilde{Y'}} 
\tilde{y} \tilde{A}^{\tilde{y}}_{-}\;) + 
 \stackrel{78}{(+)} \sum_{\{(ac)(bd) \},k,l} \; \stackrel{ac}{\tilde{(k)}} \stackrel{bd}{\tilde{(l)}}
\tilde{{A}}^{kl}_{+}((ab),(cd)) \;\;+  
  \stackrel{78}{(-)} \sum_{\{(ac)(bd) \},k,l} \; \stackrel{ac}{\tilde{(k)}}\stackrel{bd}{\tilde{(l)}}
\tilde{{A}}^{kl}_{-}((ab),(cd))\},$ (Eq.\ref{yukawa4tilde0})), does not commute with the $\Gamma^{(1,3)}$,  
they instead anticommute ($\{\Gamma^{(1,3)}, \gamma^0 \hat{m}\}_+ =0$).
But the rest of the "effective" Lagrangean (Eq.\ref{yukawa}) commutes with the operator of handedness 
in $d=(1+3)$-dimensional space: 
$\{\gamma^0\gamma^{m} (p_{m}- \sum_{A,i}\; 
g^{A}\tau^{Ai} A^{Ai}_{m}),\Gamma^{(1,3)}\}_-=0.  $

It then follows that the Lagrange density 
\begin{eqnarray}
{\mathcal L} = 
(\Gamma^{(1,3)}\psi)^{\dagger} \; \left[ \gamma^0\gamma^{m} (p_{m}- \sum_{A,i}\; 
g^{A}\tau^{Ai} A^{Ai}_{m}) - \Gamma^{(1,3)}\gamma^0 \hat{m}\Gamma^{(1,3)}\right]\; (\Gamma^{(1,3)}\psi)
\label{mto-m}
\end{eqnarray}
for the Dirac spinor $\Gamma^{(1,3)}\psi$ differs from the one from Eq.(\ref{yukawa}) 
in the sign of the 
mass term, while the function $\Gamma^{(1,3)}\psi$ differs from $\psi$ in the internal parity, if 
$\psi$ is the solution for the  Dirac equation
. Since the internal parity is just the convention, the negative mass changes sign if the 
internal parity of the spinor changes. The same argument was used in ref.(\cite{fritsch}),
while ref.(\cite{chengli}) uses the equivalent argument, namely, that the choice of the phase 
of either the right or the 
left handed spinors can always be changed and that accordingly also the signs of particular 
mass terms change. 

Let us demonstrate now on Table~\ref{TableIII} how does the operator of parity ${\cal P}$, if postulated as 
\begin{eqnarray}
\label{parity}
{\cal P} = \gamma^0 \gamma^8 I_x, {\rm with}\;\; I_x x^m (I_x)^{(-1)} =x_m,   
\end{eqnarray}
 transform a
right handed $u$-quark into the left handed $u$-quark: ${\cal P} u_{R} = \alpha u_{L}$, where 
$\alpha $ is the proportionality factor.

\begin{table}
\begin{center}
\begin{tabular}{|r|c||c||c|c||c|c|c||c|c|c||r|r|}
\hline
i&$$&$|^a\psi_i>$&$\Gamma^{(1,3)}$&$ S^{12}$&$\Gamma^{(4)}$&
$\tau^{13}$&$\tau^{21}$&$\tau^{33}$&$\tau^{38}$&$\tau^{41}$&$Y$&$Y'$\\
\hline\hline
&& ${\rm Octet},\;\Gamma^{(1,7)} =1,\;\Gamma^{(6)} = -1,$&&&&&&&&&& \\
&& ${\rm of \; quarks}$&&&&&&&&&&\\
\hline\hline
1&$u_{R}^{c1}$&$\stackrel{03}{(+i)}\stackrel{12}{(+)}|\stackrel{56}{(+)}\stackrel{78}{(+)}
||\stackrel{9 \;10}{(+)}\stackrel{11\;12}{(-)}\stackrel{13\;14}{(-)}$
&1&1/2&1&0&1/2&1/2&$1/(2\sqrt{3})$&1/6&2/3&-1/3\\
\hline 
2&$u_{R}^{c1}$&$\stackrel{03}{[-i]}\stackrel{12}{[-]}|\stackrel{56}{(+)}\stackrel{78}{(+)}
||\stackrel{9 \;10}{(+)}\stackrel{11\;12}{(-)}\stackrel{13\;14}{(-)}$
&1&-1/2&1&0&1/2&1/2&$1/(2\sqrt{3})$&1/6&2/3&-1/3\\
\hline
3&$d_{R}^{c1}$&$\stackrel{03}{(+i)}\stackrel{12}{(+)}|\stackrel{56}{[-]}\stackrel{78}{[-]}
||\stackrel{9 \;10}{(+)}\stackrel{11\;12}{(-)}\stackrel{13\;14}{(-)}$
&1&1/2&1&0&-1/2&1/2&$1/(2\sqrt{3})$&1/6&-1/3&2/3\\
\hline 
4&$d_{R}^{c1}$&$\stackrel{03}{[-i]}\stackrel{12}{[-]}|\stackrel{56}{[-]}\stackrel{78}{[-]}
||\stackrel{9 \;10}{(+)}\stackrel{11\;12}{(-)}\stackrel{13\;14}{(-)}$
&1&-1/2&1&0&-1/2&1/2&$1/(2\sqrt{3})$&1/6&-1/3&2/3\\
\hline
5&$d_{L}^{c1}$&$\stackrel{03}{[-i]}\stackrel{12}{(+)}|\stackrel{56}{[-]}\stackrel{78}{(+)}
||\stackrel{9 \;10}{(+)}\stackrel{11\;12}{(-)}\stackrel{13\;14}{(-)}$
&-1&1/2&-1&-1/2&0&1/2&$1/(2\sqrt{3})$&1/6&1/6&1/6\\
\hline
6&$d_{L}^{c1}$&$\stackrel{03}{(+i)}\stackrel{12}{[-]}|\stackrel{56}{[-]}\stackrel{78}{(+)}
||\stackrel{9 \;10}{(+)}\stackrel{11\;12}{(-)}\stackrel{13\;14}{(-)}$
&-1&-1/2&-1&-1/2&0&1/2&$1/(2\sqrt{3})$&1/6&1/6&1/6\\
\hline
7&$u_{L}^{c1}$&$\stackrel{03}{[-i]}\stackrel{12}{(+)}|\stackrel{56}{(+)}\stackrel{78}{[-]}
||\stackrel{9 \;10}{(+)}\stackrel{11\;12}{(-)}\stackrel{13\;14}{(-)}$
&-1&1/2&-1&1/2&0&1/2&$1/(2\sqrt{3})$&1/6&1/6&1/6\\
\hline
8&$u_{L}^{c1}$&$\stackrel{03}{(+i)}\stackrel{12}{[-]}|\stackrel{56}{(+)}\stackrel{78}{[-]}
||\stackrel{9 \;10}{(+)}\stackrel{11\;12}{(-)}\stackrel{13\;14}{(-)}$
&-1&-1/2&-1&1/2&0&1/2&$1/(2\sqrt{3})$&1/6&1/6&1/6\\
\hline\hline
\end{tabular}
\end{center}
\caption{\label{TableIII}%
The 8-plet of quarks - the members of $SO(1,7)$ subgroup, belonging to one Weyl left 
handed ($\Gamma^{(1,13)} = -1 = \Gamma^{(1,7)} \times \Gamma^{(6)}$) spinor representation of 
$SO(1,13)$. 
It contains the left handed weak charged quarks and the right handed weak chargeless quarks 
of a particular colour ($(1/2,1/(2\sqrt{3}))$). Here  $\Gamma^{(1,3)}$ defines the 
handedness in $(1+3)$ space, 
$ S^{12}$ defines the ordinary spin (which can also be read directly from the basic vector, 
since $S^{ab} \stackrel{ab}{(k)} = \frac{k}{2} \stackrel{ab}{(k)}$ and 
$S^{ab} \stackrel{ab}{[k]} = \frac{k}{2} \stackrel{ab}{[k]}$, if $S^{ab}$ belong to the Cartan 
subalgebra set), 
$\tau^{13}$ defines the weak charge, $\tau^{21}$ defines the $U(1)$ charge, $\tau^{33}$ and 
$\tau^{38}$ define the colour charge and $\tau^{41}$ another $U(1)$ charge, which together with the
first one defines $Y$ and $Y'$.}
\end{table}

\noindent
One notices that, if the operator ${\cal P}$ is applied on a state, which represents  
right handed weak chargeless ($\tau^{13} =0$) $u$-quark of one of the three colours  
and is presented in terms of nilpotents in the first row of Table~\ref{TableIII}, it transforms this 
state into the state, which can be found in the seventh row of the same table and represents 
the left handed $u$-quark of the same colour and spin and it is weak charged. Taking into
account Eq.(\ref{parity}) and Eq.(12,16) from ref.\cite{pikanorma05} one finds 
${\cal P} u_{R} = i u_{L}$, while 
${\cal P}{\cal P}=I$. By changing appropriately the phases of this two basic states 
($u_{R} $ and $ u_{L}$) we can easily achieve that ${\cal P} u_{R} =  u_{L},
{\cal P} u_{L} =  u_{R}$. We should in addition keep in mind that ${\cal P}$ must 
take into account also the appearance of families. We shall study discrete symmetries 
of our approach in the "low energy region" in a separate paper.


\section{Predictions with relaxed assumptions}
\label{improvedmatrices}

The mass matrices for quarks and leptons, following from the approach unifying spins and charges 
lead, after making several assumptions and simplifications 
(presented in Sects.\ref{lagrangesec} and \ref{Lwithassumptions}),   
to mass matrices with elements, which obviously very strongly correlate 
properties of the $u$-quarks, the $d$-quarks, the neutrinos and the electrons. 
The rough  estimation only makes sense, 
if the masses of the first three families are small in comparison with the masses 
of the fourth family and loses accordingly the predictive power.

We shall keep in this section some of the assumptions made in  Sects.\ref{lagrangesec} and 
\ref{Lwithassumptions} - namely the relations among the angles of rotations (
Eqs.(\ref{firstangle},\ref{phirelated},\ref{angle},\ref{abphirelated})) - 
and assume that either breaks of symmetries with some peculiar 
boundary conditions together with 
nonperturbative effects might lead to the Yukawa couplings which distinguish stronger 
among quarks and leptons than seen above.

We keep the following assumptions: i)  {\em matrices are symmetric 
and real}, ii) {\em diagonalization in two steps} (Eq.\ref{deggen}) {\em 
is possible},  and iii)  the  
{\em relations among the angles of rotations for the $u$-quarks  and  the $d$-quarks 
matrices, as well as among the angles of rotations 
for the matrices of the neutrinos and the electrons,   
which determine the first and the second step diagonalizations, stay related as presented in} (Sect.
\ref{Lwithassumptions})  Eqs.(\ref{phirelated},\ref{abphirelated}). 

We {\em assume that fields} $\tilde{\omega}_{abc}$ on Table~\ref{TableI} and~\ref{TableII} (Sect.\ref{lagrangesec}) 
{\em carry an }(additional) {\em index} $\alpha$ ($\tilde{\omega}_{abc\alpha}$), {\em which 
distinguishes among} $u$-{\em quarks}, $d$-{\em quarks}, {\em neutrinos and electrons.} 
Any break of symmetries would further relate $\tilde{\omega}_{abc\alpha}$ but might also 
make differences in the properties of  the members of one family more expressed.

We hope that when trying to reproduce the experimental data, the ratios among the fields 
$\tilde{\omega}_{abc \alpha}$ will tell us something about the 
 break of symmetries or about other possible reasons for so different 
 properties of quarks and leptons even on this very preliminary stage of studying the 
 predictions of the approach unifying spins and charges.

 It follows then that in Eqs.(\ref{aunuIIa},\ref{aunuIIb},%
\ref{adeIIa},\ref{adeIIb}) all the fields carry an additional  index $\alpha = u,\nu,d,e,$ 
 while we keep from the previous study  the relations $k_{\alpha} = - k_{\beta}$, 
$\alpha = u, \nu$ and $\beta = d, e$, where $k_{\alpha,\beta}$ define the first step 
orthogonal transformations leading to $2$ by $2$ by diagonal mass matrices and  
the relations among the angles of rotations in the second step of 
orthogonal transformations determined by ${}^{a,b}\eta_{\alpha} = {}^{a,b}(\frac{2B}{C-A})_{\alpha}$, 
requiring that (Eq.(\ref{abphirelated})) 
${}^{a,b}\eta_{\alpha}=  -{}^{a,b}\eta_{\beta},$
where $C,A,B$ are replaced 
by the corresponding matrix elements for the first two families determined by the matrix 
$A^a_{\alpha,\beta}$ (Eqs.(\ref{aunuIIa}, \ref{adeIIa})) 
and the second two families determined by the matrix $A^b_{\alpha,\beta}$ (%
Eqs.(\ref{aunuIIb}, \ref{adeIIb})).

It then follows  
\begin{eqnarray}
\label{angleab}
{}^a\varepsilon_{\alpha}\;(\tilde{\omega}_{018 \beta}- 
\sqrt{1+ (\frac{k_{\alpha}}{2})^2 }\; \tilde{\omega}_{187 \beta}) &=&
 ( \tilde{\omega}_{018 \alpha}- \sqrt{1+ (\frac{k_{\alpha}}{2})^2 }\; 
\tilde{\omega}_{187 \alpha}),\nonumber\\
{}^b\varepsilon_{\alpha}\; (\tilde{\omega}_{018 \beta}+ \sqrt{1+ (\frac{k_{\alpha}}{2})^2 } \;
\tilde{\omega}_{187 \beta}) &=&
( \tilde{\omega}_{018 \alpha} + \sqrt{1+ (\frac{k_{\alpha}}{2})^2 } \;
\tilde{\omega}_{187 \alpha}),\nonumber\\
{}^a\varepsilon_{\alpha}\;(\tilde{\omega}_{127 \beta}+ \sqrt{1+ (\frac{k_{\alpha}}{2})^2 } \;
\tilde{\omega}_{078 \beta}) &=&
 ( \tilde{\omega}_{127 \alpha}+ \sqrt{1+ (\frac{k_{\alpha}}{2})^2 } \;
\tilde{\omega}_{078 \alpha}),\nonumber\\
{}^b\varepsilon_{\alpha}\;(\tilde{\omega}_{127 \beta}- \sqrt{1+ (\frac{k_{\alpha}}{2})^2 } \;
\tilde{\omega}_{078 \beta}) &=&
 ( \tilde{\omega}_{127 \alpha}- \sqrt{1+ (\frac{k_{\alpha}}{2})^2 } \;
\tilde{\omega}_{078 \alpha}),
\end{eqnarray}
where index $a$ and $b$ distinguish between the two by two matrices for the first two and 
the second two families, correspondingly, while $\alpha =u,\nu$ and $\beta=d,e$.

The question is whether or   not  the three quantities $k_{\alpha}, 
{}^a\eta_{\alpha}, {}^b\eta_{\alpha}$, $\alpha = u, \nu,$ for the quarks and the three 
 for the leptons are enough to fit the existing experimental data  
 for the quark and the lepton mixing matrices. Both have namely the form 
\begin{eqnarray}
\label{abcwithm}
{\bf V}_{\alpha \beta} = \left(\begin{array}{cccc}
c(\varphi)c({}^a\varphi)&-c(\varphi)s({}^a\varphi)&
-s(\varphi)c({}^a\varphi^b)& s(\varphi)s({}^a\varphi^b)\\
c(\varphi)s({}^a\varphi)&c(\varphi)c({}^a\varphi)&
-s(\varphi)s({}^a\varphi^b)& -s(\varphi)c({}^a\varphi^b)\\
s(\varphi)c({}^a\varphi^b)&-s(\varphi)s({}^a\varphi^b)&
c(\varphi)c({}^b\varphi)& -c(\varphi)s({}^b\varphi)\\
s(\varphi)s({}^a\varphi^b)&s(\varphi)c({}^a\varphi^b)&
c(\varphi)s({}^b\varphi)& c(\varphi)c({}^a\varphi)\\
 \end{array}
                \right),
\end{eqnarray}
with the angles (Eq.\ref{phirelated},\ref{abphirelated}) described  by the three parameters 
$k_{\alpha},{}^{a}\eta_{\alpha},{}^{b}\eta_{\alpha}$ 
as follows
\begin{eqnarray}
\label{anglesandmixmatr3ngles}
\varphi = \varphi_{\alpha}-\varphi_{\beta},
\quad {}^a\varphi = {}^a\varphi_{\alpha}-{}^a\varphi_{\beta},\quad 
{}^a\varphi^b = - \frac{{}^a\varphi + {}^b \varphi}{2}.
\end{eqnarray}
If the mixing matrix for either quarks or leptons is described by the three parameters (each)
$k_{\alpha}, {}^a\eta_{\alpha}, {}^b\eta_{\alpha}; \alpha=u,\nu,$, then we have just 
enough free parameters  to make any choice for the masses of the fourth family of quarks and leptons.   
To see this we just express $A,B,C$ in any of the two $2$ by $2$ 
matrices in terms of the corresponding diagonal values that is in terms of the masses $m_{\alpha i}, 
m_{\beta i};
i=1,2,3,4; \alpha= u,\nu,\beta=d,e,$ and the parameters 
$k_{\alpha}=-k_{\beta}, {}^a\eta_{\alpha} = - {}^a\eta_{\beta}, {}^b\eta_{\alpha} = - {}^b\eta_{\beta};
\alpha=u,\nu,$.
The matrix elements of ${\bf A}^{a}_{\alpha}$ (${}^aa_{\alpha},{}^ab_{\alpha},{}^ac_{\alpha}
$), for $u$-quarks and neutrinos are expressible with 
the masses $m_{\alpha 1}, m_{\alpha 2}$ of the first two families 
of $u$-quarks or neutrinos and the corresponding angles of 
rotations as follows
\begin{eqnarray}
\label{abcwithm}
{\bf A}^{a}_{\alpha} = \left(\begin{array}{cc}
\frac{1}{2}( m_{\alpha 1} + m_{\alpha 2} - \frac{{}^a\eta_{\alpha} (m_{\alpha 2} - 
m_{\alpha 1})}{\sqrt{1+({}^a\eta_{\alpha})^2}}),& 
\frac{m_{\alpha 2} - m_{\alpha 1}}{2\sqrt{1+({}^a\eta_{\alpha})^2}})\nonumber\\
\frac{m_{\alpha 2} - m_{\alpha 1}}{2\sqrt{1+({}^a\eta_{\alpha})^2}}),&
\frac{1}{2}( m_{\alpha 1} + m_{\alpha 2} + 
\frac{{}^a\eta_{\alpha}(m_{\alpha 2} - m_{\alpha 1})}{\sqrt{1+({}^a\eta_{\alpha})^2}})
 \end{array}
                \right),
\end{eqnarray}
while the 
expressions for the matrix ${\bf A}^{b}_{\alpha}$ with matrix elements 
${}^ba_{\alpha}, {}^bc_{\alpha}, {}^bb_{\alpha}$ follow, if
we replace $m_{\alpha 1}$ with $m_{\alpha 3}$ and $m_{\alpha 2}$ with $m_{\alpha 4}$.
Equivalently we obtain the mass matrices for the $d$-quarks and the 
electrons by replacing $\alpha$ by $\beta$ in all expressions. 
Eq.(\ref{abcwithm}) below demonstrates that if once  the three parameters 
$k_{\alpha},{}^{a}\eta_{\alpha},{}^{b}\eta_{\alpha}$ are chosen to fit the
experimental data, any four masses for the fourth family of quarks and leptons 
agree with the proposed requirements.

The starting mass matrices ${\bf M}_{\alpha,\beta} $ - the Yukawa couplings - for 
quarks and leptons, 
which are $4$ by $4$ matrices and are the generalized versions of 
Table~\ref{TableI} and Table~\ref{TableII}, are expressible 
with the matrices  ${\bf A}^{a,b}_{\alpha,\beta}$ of Eq.(\ref{abcwithm}) as follows
\begin{eqnarray}
\label{yukamainm}
 \left(\begin{array}{cc}
\frac{1}{2}[({\bf A}^a_{\alpha,\beta } + {\bf A}^b_{\alpha,\beta }) - ({\bf A}^b_{\alpha,\beta } - 
{\bf A}^a_{\alpha,\beta })\;
\frac{k_{\alpha,\beta}}{2\sqrt{1+(\frac{k_{\alpha,\beta}}{2})^2}}],& 
 ({\bf A}^b_{\alpha,\beta} - {\bf A}^a_{\alpha,\beta}) \frac{1}{2\sqrt{1+
 (\frac{k_{\alpha,\beta}}{2})^2}}) \nonumber\\
({\bf A}^b_{\alpha,\beta } - {\bf A}^a_{\alpha,\beta }) 
\frac{1}{2\sqrt{1+(\frac{k_{\alpha,\beta}}{2})^2}}),&
\frac{1}{2}[({\bf A}^a_{\alpha,\beta} + {\bf A}^b_{\alpha,\beta}) + ({\bf A}^b_{\alpha,\beta} - 
{\bf A}^a_{\alpha,\beta})\;
\frac{k_{\alpha,\beta}}{2\sqrt{1+(\frac{k_{\alpha,\beta}}{2})^2}}]
 \end{array}
                \right).
\end{eqnarray}
One easily sees that the matrix 
${\bf M_{\alpha,\beta}}$ is 
equal to a democratic matrix with all the elements equal 
to $m_{\alpha_4}/4, $ with $\alpha = u, \nu, d, e,$ if all the angles of rotations are equal to $\pi/4$   
(that is for $k_{\alpha}=0$, ${}^{a,b}\eta_{\alpha} = 0$), while 
$m_{\alpha_i,\beta_i}=0, i=1,3$,  and that the mixing matrices are 
then the identity.

Once knowing the matrices ${\bf M_{\alpha,\beta}} $, one easily derives the parameters 
 $\tilde{\omega}_{abc\alpha,\beta},$ with $(abc)$ equal to $(018),
(078), (127),(187),(387)$, which (in our generalized version) enter into Table~\ref{TableI} 
and Table~\ref{TableII}. 
One namely finds
\begin{eqnarray}
\label{omegatilde}
\tilde{\omega}_{018\alpha} &=&\frac{1}{2}[\frac{m_{\alpha 2} - 
m_{\alpha 1}}{\sqrt{1+ ({}^a\eta_{\alpha})^2}} +
\frac{m_{\alpha 4 } - m_{\alpha 3}}{ \sqrt{1+ ({}^b\eta_{\alpha})^2}}], \nonumber\\
\tilde{\omega}_{078\alpha} &=&\frac{1}{2 \;\sqrt{1+ (\frac{k_{\alpha}}{2})^2}}
[\frac{{}^a\eta_{\alpha}\;(m_{\alpha 2} - m_{\alpha 1})}{\sqrt{1+ ({}^a\eta_{\alpha})^2}} - 
\frac{{}^b\eta_{\alpha}\;(m_{\alpha 4 } - m_{\alpha 3})}{\sqrt{1+ ({}^b\eta_{\alpha})^2}}], \nonumber\\
 \tilde{\omega}_{127\alpha} &=&\frac{1}{2}[\frac{{}^a\eta_{\alpha}\;(m_{\alpha 2} - 
m_{\alpha 1})}{\sqrt{1+ ({}^a\eta_{\alpha})^2}} +
\frac{{}^b\eta_{\alpha}\;(m_{\alpha 4 } - m_{\alpha 3})}{\sqrt{1+ ({}^b\eta_{\alpha})^2}}], \nonumber\\
\tilde{\omega}_{187\alpha} &=&\frac{1}{2 \sqrt{1+(\frac{k_{\alpha}}{2})^2}}[\frac{-
m_{\alpha 2} - m_{\alpha 1}}{\sqrt{1+ ({}^a\eta_{\alpha})^2}} +
\frac{m_{\alpha 4 } - m_{\alpha 3}}{ \sqrt{1+ 
({}^b\eta_{\alpha})^2}}], \nonumber\\
\tilde{\omega}_{387\alpha} &=& \frac{1}{2 \sqrt{1+(\frac{k_{\alpha}}{2})^2}}
[(m_{\alpha 4 } + m_{\alpha 3}) - (m_{\alpha 2} + m_{\alpha 1})],\nonumber\\
a^a_{\alpha}&=& \frac{1}{2}( m_{\alpha 1} + m_{\alpha 2} - \frac{{}^a\eta_{\alpha}\;(m_{\alpha 2} - 
m_{\alpha 1})}{\sqrt{1+({}^a\eta_{\alpha})^2}}).
\end{eqnarray}
%


\section{Numerical results}
\label{numerical}

In this section we connect parameters  $\tilde{\omega}_{abc}$ of the Yukawa couplings 
following from our approach unifying spins and charges (after several assumptions, approximations 
and simplifications) with the experimental data. We investigate, how the parameters of the 
approach reflect the known data. We also investigate a possibility of making some predictions.


\subsection{Experimental data for quarks and leptons}
\label{experimentaldata}

We present in this subsection those experimental data, which are relevant for our study:  
that is the measured values for the masses of the three 
families of quarks and leptons and the measured mixing matrices.

We take in our calculations the experimental masses for the known three 
families from the ref.\cite{expckm}.
\begin{eqnarray}
\label{masses}
m_{u_i}/GeV &=& (0.0015- 0.004, 1.15-1.35, 174.3-178.1
),\nonumber\\
m_{d_i}/GeV &=& (0.004-0.008, 0.08-0.13, 4.1-4.9 
),\nonumber\\
m_{\nu_i}/GeV &=& (1 \; 10^{-12}, 1 \; 10^{-11}, 5 \; 10^{-11} 
),\nonumber\\
m_{e_i}/GeV &=& (0.0005,0.105,1.8
).
\end{eqnarray}
Predicting four families of quarks and leptons at ''physical'' energies, we require 
the unitarity condition for the mixing matrices for four rather than three measured families 
of quarks\cite{expckm}  
\begin{eqnarray}
\label{expckm}
 \left(\begin{array}{ccc}
 0.9730-0.9746 & 0.2174-0.2241 & 0.0030-0.0044\\
 0.213-0.226   & 0.968-0.975  & 0.039-0.044\\
 0.0 - 0.08    & 0.0-0.11      & 0.07-0.9993\\
\end{array}
                \right).
\end{eqnarray}
The experimental data are for the mixing matrix for leptons known very 
weakly\cite{expmixleptons} 
\begin{eqnarray}
\label{expckmleptons}
 \left(\begin{array}{ccc}
 0.79-0.88 & 0.47-0.61 &  {} < 0.20\\
 0.19-0.52 & 0.42-0.73 & 0.58-0.82\\
 0.20-0.53 & 0.44-0.74 & 0.56-0.81\\
\end{array}
                \right).
\end{eqnarray}
We see that within the experimental accuracy both mixing matrices - for  
quarks and leptons - may be assumed to be symmetric up to a sign. We then fit with these 
two matrices the six parameters $k_{\alpha}, {}^a\eta_{\alpha}, {}^b\eta_{\alpha}$, $\alpha =u,\nu.$ 


\subsection{Resuls}
\label{numericalresults}

We started with the explicit expressions for the Yukawa couplings suggested by the approach 
unifying spins and charges and   made several assumptions 
and approximations, also simplifications, in order to be able to make some approximate  
predictions.  
Since we do not know the way of breaking symmetries for either the Poincar\' e group or for 
the group defining families  and how breaking of these two kinds of symmetries is connected  
and to which properties of the system would they 
lead - and whether they would or not accordingly support the assumptions and 
approximations we made - we proceeded in two steps. The first step brought us to  strongly related 
mass matrices for quarks and leptons, suggesting that the fourth family of quarks and leptons 
lies very high. In the second step we keep those symmetries of the mass matrices  
suggested by the first step, 
which lead to correlated rotations  for the $u$-quarks and  the $d$-quarks on one side 
and for the neutrinos and the electrons on the other side, but allow that  
$\tilde{\omega}_{abc}$ fields 
might not be the same for the $u$ and $\nu$ and $d$ and $e$, hoping  that  a kind of 
breaking symmetries with special boundary conditions and nonperturbative effects  
might effect these fields in the assumed way.

We can now  connect the parameters of the approach 
(left after several approximations and assumptions) with the experimental data and  
try to find out what can we learn from the corresponding results. As we have said: 
Any choice for the masses of the fourth family fits the experimental data, once 
(twice) the three angles of the orthonormal transformations, determining the (two) mixing
matrices are chosen. 

{\em We fit} (twice) {\em the three angles of Eqs.(\ref{phirelated},\ref{abphirelated}) 
with the Monte-Carlo method under the requirement that the ratios of the parameters 
$\tilde{\omega}_{abc}$, entering into mass matrices, are so close to a rational number as possible.}
This requirement is made in order to see whether some kind of symmetry might be 
responsible for the difference 
in properties of quarks and leptons. 

We allow the masses of the fourth family as follows: 
The two quark masses must lie in the range from $200$ GeV to $1$ TeV, the fourth neutrino mass 
must be within the interval $50 - 100$ GeV and of the fourth electron mass within $50 - 200$ GeV. 

\begin{figure}
\includegraphics[width=8cm]{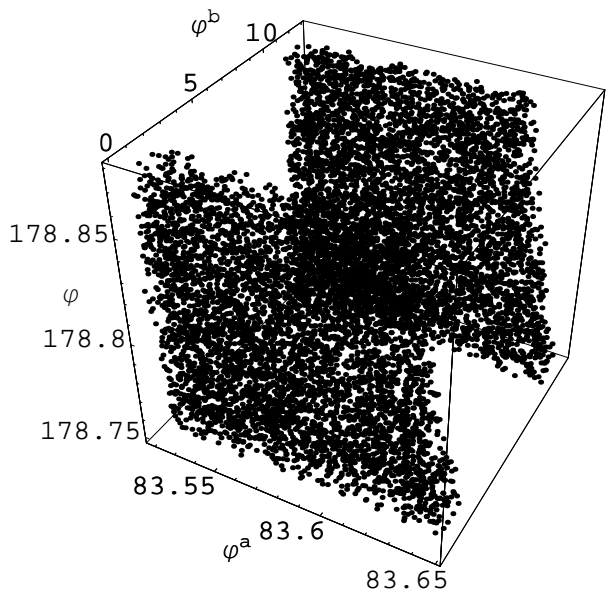}
\caption{\label{fig1}%
Figure shows the Monte-Carlo fit\cite{matjazdiploma} of the experimental mixing matrix 
for quarks (Eq.\ref{expckm}) with the three angles of Eq.(\ref{abcwithm}). The three angles  
define the three parameters $k_u, {}^a\eta_u$ and ${}^b\eta_u$ 
(Eqs.\ref{firstangle},\ref{solutions}). We make a choice among those values for the best  fit, 
 which makes the ratios $\tilde{\omega}_{abc u}/ \tilde{\omega}_{abc d}$ as close to 
 rational numbers as possible while assuring that the masses of the three known families stay 
 within the   acceptable values from Eq.(\ref{masses}), with no constraints on $a_{\alpha}$ and 
 the two quark masses of the fourth family lie in the range $200-1000$ MeV.}
\end{figure}

Fig.~\ref{fig1} shows the results of the Monte-Carlo simulation for the three angles determining the 
mixing matrix for quarks. There are the  experimental inaccuracies, which determine the allowed regions 
for the three angles. 

The results for the quarks are presented on Table~\ref{TableIV} and~\ref{TableV} (together with 
the corresponding values for leptons).

\begin{figure}
\includegraphics[width=8cm]{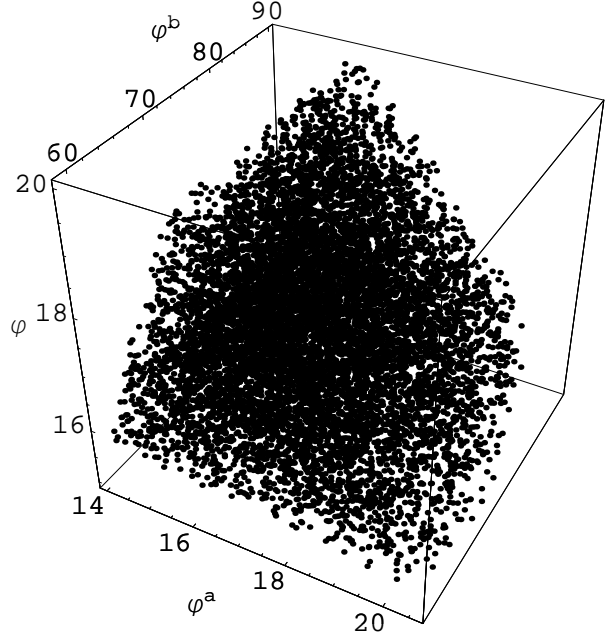}
\caption{\label{fig2}%
Figure shows  the Monte-Carlo fit of the experimental data for the mixing matrix for 
leptons(Eq.(\ref{expckmleptons})).  The three angles  
define the three parameters $k_{\nu}, {}^a\eta_{\nu}$ and ${}^b\eta_{\nu}$ 
(Eqs.\ref{firstangle},\ref{solutions}). Again we make a choice among those values for the best  fit, 
 which make the ratios $\tilde{\omega}_{abc u}/ \tilde{\omega}_{abc d}$ as close to 
 rational numbers as possible. }
\end{figure}

Fig.~\ref{fig2} shows the Monte-Carlo fit for the three angles determining the mixing matrix for
leptons. There are the experimental inaccuracies which limit the values of the three angles. 
Again we make a choice among those values for the best  fit, 
 which make the ratios $\tilde{\omega}_{abc u}/ \tilde{\omega}_{abc d}$ as close to 
 rational numbers as possible. Since in the lepton case the mixing matrix for the 
 three known families as well as the masses for the three  neutrinos  are weakly known, the 
 calculations for four families bring much less information than in the quark case.

The results for the leptons are presented together with the results for the quarks 
 on Table~\ref{TableIV} and~\ref{TableV}.

\begin{table}
\centering
\begin{tabular}{|c||c|c|c|c|}
\hline 
&$u$&$d$&$\nu$&$e$\tabularnewline
\hline
\hline 
$k$       & -0.085 &  0.085&-1.254&  1.254\tabularnewline
\hline 
${}^a\eta$& -0.229 &  0.229& 1.584& -1.584\tabularnewline
\hline
${}^b\eta$&  0.420 & -0.440& -0.162& 0.162\tabularnewline
\hline
\end{tabular}\\
\caption{\label{TableIV}%
The Monte-Carlo fit to the experimental data\cite{expckm,expmixleptons} 
for the three parameters $k$, ${}^a\eta$ and   ${}^b\eta$ determining the mixing matrices 
for  the four families of quarks and leptons are presented.}
\end{table}

\begin{table}
\begin{tabular}{|c||c|c|c||c|c|c|}
\hline 
&$u$&$d$&$u/d$&$\nu$&$e$&$\nu/e$\tabularnewline
\hline
\hline 
$|\tilde{\omega}_{018}|$& 21205& 42547& 0.498&    10729& 21343 &0.503\tabularnewline
\hline 
$|\tilde{\omega}_{078}|$& 49536& 101042& 0.490&   31846& 63201 &0.504\tabularnewline
\hline 
$|\tilde{\omega}_{127}|$& 50700& 101239& 0.501&   37489& 74461 &0.503\tabularnewline
\hline 
$|\tilde{\omega}_{187}|$& 20930& 42485& 0.493&    9113&  18075 &0.505\tabularnewline
\hline 
$|\tilde{\omega}_{387}|$& 230055& 114042& 2.017&  33124& 67229 &0.493\tabularnewline
\hline
$a^{a}$&94174& 6237& &   1149&1142 &\tabularnewline
\hline
\end{tabular}\\
\caption{\label{TableV}%
Values for the parameters $\tilde{\omega}_{abc}$ (entering into 
the mass matrices for the $u-$quarks, the $d-$quarks, the neutrinos and the 
electrons, as suggested by the approach) as following after the Monte-Carlo fit, 
relating the parameters and the experimental data.}
\end{table}

In Eq.(\ref{resultmasses}) we present masses for the four families of quarks and leptons as 
obtained after the Monte-Carlo fit
\begin{eqnarray}
\label{resultmasses}
m_{u_i}/GeV &=& (0.0034, 1.15, 176.5, 285.2),\nonumber\\
m_{d_i}/GeV &=& (0.0046, 0.11, 4.4, 224.0), \nonumber\\
m_{\nu_i}/GeV &=& ( 1\; 10^{-12}, 1 \; 10^{-11}, 5 \; 10^{-11},  84.0 ),\nonumber\\
m_{e_i}/GeV &=& (0.0005,0.106,1.8, 169.2).
\end{eqnarray}

The results of the Monte-Carlo fit shows that the requirement, that the ratios of the 
corresponding parameters of $\tilde{\omega}_{abc}$ for the quarks and the leptons  should 
be as close to the rational numbers as possible, makes that the fourth 
family lies within the experimentally allowed values as evaluated by the 
refs.\cite{okun,okunmaltoni,okunbulatov}. 
Eq.(\ref{omegatilde}), however, tells us, that it is the top mass and 
the masses of the fourth family which mostly (not entirely) determine these ratios. 
But integer or half integer ratios could still be 
a sign that some group properties or even nonperturbative effects connected with the 
charges  of quarks and leptons determine the masses of fermions, since if we move the masses from 
those allowed by the refs.\cite{okun,okunmaltoni,okunbulatov}, the ratios go to one 
only when all the masses of the fourth family are equal and 
are high in comparison with the top mass.

The Monte-Carlo fit  leads to the following mixing matrix for the quarks
\begin{eqnarray}
\label{resultckm}
 \left(\begin{array}{cccc}
 0.974 & 0.223 & 0.004 & 0.042\\
 0.223 & 0.974 & 0.042 & 0.004\\
 0.004 & 0.042 & 0.921 & 0.387\\
 0.042 & 0.004 & 0.387 & 0.921\\
 \end{array}
                \right)
\end{eqnarray}
and for the leptons
\begin{eqnarray}
\label{resultckmleptons}
 \left(\begin{array}{cccc}
 0.697 & 0.486 & 0.177 & 0.497\\
 0.486 & 0.697 & 0.497 & 0.177\\
 0.177 & 0.497 & 0.817 & 0.234\\
0.497  & 0.177 & 0.234 & 0.817\\ 
\end{array}
                \right).
\end{eqnarray}
The estimated mixing matrix for the four families of quarks predicts quite a strong couplings  
between the fourth and the other three families, limiting  (due to the assumptions and 
approximations we made, which manifest in the symmetric mixing matrices) some of the matrix elements 
of the three families as well.  

The estimated mixing matrix for the four families of leptons predicts very probably far 
too strong couplings 
between the known three and the fourth family (although they are not in contradiction  with  
the report in\cite{expckm}).

Let us end  up this section by repeating that all the predictions must be taken as  a very rough 
estimation, since they follow from the approach unifying spins and charges after  many 
approximations and assumptions, which
we made to be able to come in quite a short way to simple and transparent predictions.


\section{Discussions and conclusions}
\label{discussions}

In this paper and in the previous one\cite{pikanorma05} we study a possibility 
that the approach of one of 
us\cite{norma92,norma93,normasuper94,norma95,norma97,pikanormaproceedings1,holgernorma00,norma01,%
pikanormaproceedings2,Portoroz03}, unifying spins and charges, might be a new right way 
for answering those of 
the open questions of the Standard model of the electroweak and colour interaction, 
which are connected with the appearance of  families of fermions, of the Yukawa couplings and of  
the weak scale: Why do only the left handed spinors carry the weak 
charge, while the right handed are weak chargeless? 
Where do the families of the quarks and the leptons come from? 
What does determine the strenghts of the Yukawa 
couplings and the weak scale?

Within the approach unifying spins and charges the answer to  the question, why 
only the left handed spinors carry the weak charge, while the right handed are 
weak chargeless, does exist: The representation of one Weyl spinor of the group SO(1,13), 
analyzed with respect to the properties of the subgroups SO(1,7)x SU(3)xU(1) of this 
group and further with respect to SU(2) and the second U(1), manifests the left 
handed weak charged quarks and leptons and the right handed weak chargeless quarks and 
leptons. 

The approach answers as well the question about a possible origin of the  ''dressing'' of the right 
handed quarks and leptons in the  Standard model: The approach proposes 
the Lagrange density for fermions in $d(=1+13)$-dimensional space in which the gauge 
field of the Poincar\' e group is the only interaction through spin connections and vielbeins. 
It is  a part of the spin connection field, which connects the right handed weak 
chargeless spinors with the left handed weak charged ones, playing the role of the Higgs field 
(and the Yukawa couplings within a family) of the Standard model. 

The approach is answering also the question about the origin of the families of 
quarks and leptons: Two kinds of the Clifford algebra objects gauging 
two kinds of the spin connection fields, are asumed. One kind 
takes care of the spin and the charges and of connecting right handed weak chargeless 
fermions with left handed weak charged fermions. The other kind takes care 
of  the families of fermions and consequently 
of the Yukawa couplings among the families contributing also to the diagonal elements. 
In the previous paper\cite{pikanorma05} we derived from the approach - after making 
several approximations, assumptions and simplifications - the expressions for the Yukawa couplings 
for four families of quarks and leptons. Approximations, assumptions and simplifications lead  
to very simple expressions for the mass matrices for the four families of quarks and leptons 
in terms of the spin connection fields of the two kinds. 

The approximate break of the symmetry - from $SO(1+5)$ to $SU(2)\times
SU(2)\times U(1)$ in the $\tilde{S}$ sector - 
suggests that three angles might in quite a good approximation determine the 
mixing matrices for the four families of quarks and leptons. We use this suggestion 
to simplify further estimations. We must, however,  add that an approximate 
break of the symmetry 
from $SO(1+5)$ to $SU(3)\times U(1)$ instead would suggest that the fourth family is 
very weakly coupled 
to the first three and would accordingly strongly change our - very preliminary - 
results. (While such a break seems to  be even acceptable when describing properties of leptons, 
it would  predict for quarks much too strong couplings between the third and the first two families  
than they measured.) 

Not knowing the way of breaking symmetries from $SO(1,13)$ to the observed ones for any of 
the two types of the symmetries (the Poincar\' e one and the one connected with the generators, 
$\tilde{S}^{ab}$), we could only guess it through assumptions which do not contradict 
the experimental data and by treating the breaking in both sectors ($S^{ab}$ and $\tilde{S}^{ab}$) 
equivalently as much as possible. We assume that effects like the breaking of 
symmetries or nonperturbative effects 
might be responsible 
for the difference in the nondiagonal matrix elements of the Yukawa couplings, while in the 
diagonal ones the difference in matrix elements originates also in the difference in the 
quantum numbers carried by quarks and leptons. 

We treat quarks and leptons equivalently and did not take into account a possible 
existence of the Majorana neutrinos: 
all the masses are the Dirac masses. 

We make in this paper a rough prediction of the  properties of the fourth family for quarks and leptons 
by connecting the parameter of our approach with the experimental data. We fix  the masses 
of the fourth family by requiring that the ratios of the corresponding parameters of the approach 
for quarks and leptons are as close to the rational numbers as possible. We get numbers like 
$\frac{1}{2}$ or $ 1$ for these ratios and let for further studies to better understand 
the influence of the way of breaking symmetries and of the nonperturbative (or perturbative) 
effects on the properties of families at "physical energies".

Our rough estimation of the properties of the fourth family agrees with the analyses of 
refs.\cite{okun,okunmaltoni,okunbulatov} and it predicts the fourth family masses $m_{u_4}=285$ GeV,
$m_{d_4}=224$ GeV, $m_{\nu_4}=65$ GeV, $m_{e_4}=129$ GeV. The mixing matrices are  in our rough 
prediction symmetric since mass matrices are assumed to be symmetric and real. 
Predictions for the couplings between the fourth and the other three families seem reasonable for quarks, 
while for leptons the corresponding mixing matrix elements might suggest  that either different break of 
symmetries in the 
$\tilde{S}^{ab}$ sector from the assumed one, or the Majorana neutrinos, or both effects should at least be 
further studied.

To try to answer within the approach unifying spins and charges the open question of the 
Standard model: Why the weak scale appears as it does? a more detailed study of the breaks 
of symmetries in both sectors is needed.

\section*{Acknowledgments} We would like to express many thanks to ARRS for the
grant.
It is a pleasure to thank all the participants of the   workshops entitled 
"What comes beyond the Standard model", 
taking place  at Bled annually in  July, starting at 1998,  for fruitful discussions, 
in particular to H.B. Nielsen.


 \end{document}